\documentclass[conference]{IEEEtran}

\usepackage[left=1.62cm,right=1.62cm,top=2.7cm]{geometry}

\usepackage{cite}
\usepackage[english]{babel}
\usepackage[utf8]{inputenc}
\usepackage[noend]{algpseudocode}
\usepackage{amsmath,amssymb,amsfonts}
\usepackage{verbatim}
\usepackage{graphicx}
\usepackage{textcomp}
\usepackage{float}
\usepackage{caption}
\usepackage[font=small]{caption}
\usepackage{soul,xcolor}
\usepackage{comment}
\usepackage{multirow}
\usepackage{color, colortbl}
\usepackage[ruled,vlined]{algorithm2e}

\usepackage{textcomp,amssymb}
\usepackage{setspace}
\usepackage{latexsym,fancyhdr,url}
\usepackage{enumerate}
\usepackage{adjustbox}
\usepackage[ruled,vlined]{algorithm2e}
\usepackage{import}
\usepackage{xcolor}
\usepackage{color, soul}
\usepackage{subfig}

\begin{document}

\IEEEoverridecommandlockouts
\IEEEpubid{\makebox[\columnwidth]{978-1-5708-0592-5/22/\$31.00~\copyright2022 IEEE \hfill} \hspace{\columnsep}\makebox[\columnwidth]{ }}

\title{Mitigation of Cyberattacks through Battery Storage for Stable Microgrid Operation}

\author{
\IEEEauthorblockN{\textbf{Ioannis Zografopoulos}\IEEEauthorrefmark{1}, \textbf{Panagiotis Karamichailidis}\IEEEauthorrefmark{1}, \textbf{Andreas T. Procopiou}\IEEEauthorrefmark{2}, 
\textbf{Fei Teng}\IEEEauthorrefmark{3}, \\ \textbf{George C. Konstantopoulos}\IEEEauthorrefmark{4}, 
\textbf{Charalambos Konstantinou}\IEEEauthorrefmark{1}} \\
\IEEEauthorblockA{
\IEEEauthorrefmark{1}CEMSE Division, King Abdullah University of Science and Technology (KAUST)\\
\IEEEauthorrefmark{2}Watts Battery Corp., Palo Alto, CA \\
\IEEEauthorrefmark{3}Dept. of Electrical and Electronic Engineering, Imperial College London\\
\IEEEauthorrefmark{4}Dept. of Electrical and Computer Engineering, University of Patras\\
E-mail: \{ioannis.zografopoulos, panagiotis.karamichailidis, charalambos.konstantinou\}@kaust.edu.sa,\\ andreasprocopiou@ieee.org, f.teng@imperial.ac.uk, g.konstantopoulos@ece.upatras.gr} \\ 
}
\IEEEaftertitletext{\vspace{-1\baselineskip}}

\maketitle

\begin{abstract}
In this paper, we present a mitigation methodology that leverages battery energy storage system (BESS) resources in coordination with microgrid (MG) ancillary services to maintain power system operations during cyberattacks. The control of MG agents is achieved in a distributed fashion, and once a misbehaving agent is detected, the MG's mode supervisory controller (MSC) isolates the compromised agent and initiates self-healing procedures to support the power demand and restore the compromised agent. Our results demonstrate the practicality of the proposed attack mitigation strategy and how grid resilience can be improved using BESS synergies. Simulations are performed on a modified version of the Canadian urban benchmark distribution model. 
\end{abstract}

\begin{IEEEkeywords}
Cyber-physical microgrids, battery energy storage systems, attacks, detection, mitigation.
\end{IEEEkeywords}

\section{Introduction} \label{s:Introduction}

In recent decades, the power grid has undergone significant overhauls. Microgrids (MGs), demand response schemes, day-ahead markets, etc.  support the provision of flexible energy to consumers, overcoming the dependence on centrally generated energy from geographically dispersed facilities. Inverter-based resources (IBRs) and battery energy storage systems (BESS) contribute significantly to this end, while alleviating power consumption ``hotspots'' supporting peak-shaving mechanisms.

As information and communication technologies are incorporated into power systems to harness the advantages of the deployed distributed generations (DGs), the threat landscape of such cyber-physical energy systems (CPES) is expanded. Thus, it is crucial to realize security mechanisms that can effectively detect malicious grid behavior and leverage the power grid's resources to overcome such adverse phenomena. Diverse cyberattacks targeting different layers of CPES have been explored in  literature \cite{zografopoulos2021cyber}. 
For instance, attacks could target the communication between DGs, delaying or modifying packages, thus causing denial-of-service (DoS) attacks \cite{zografopoulos2020derauth}. Attackers could also perform firmware modification attacks on IBRs that lead to uneconomical system operation or even trigger blackouts and jeopardize grid equipment \cite{zografopoulos2022time}.

To harden CPES against adversaries, compound security and intrusion-tolerant mechanisms, that can detect anomalous behavior and proactively perform remediation actions, leveraging system resources need to be automated. {For example, in \cite{zografopoulos2021detection}, the authors propose the use of matrix subspace methods for the detection of cyberattacks targeting autonomously operating MGs and present the potential impact of such attacks if not immediately identified.} The importance of system resources, such as BESS, in supporting ancillary services of the grid is discussed in \cite{li2019review}. Different scenarios are presented where BESS can be used to minimize power fluctuations, regulate grid voltage and frequency, 
and support distributed systems under cyberattacks or other unexpected conditions. In \cite{6194234}, the authors present an algorithm that can coordinate the contribution of distributed energy resources (DERs), in terms of active and reactive power, thus enhancing the stability of the system. 

Control strategies have also been employed to maintain security and resiliency in CPES \cite{konstantinou2021resilient}. For instance, a 
control mechanism, leveraging power sharing between multiple energy storage resources, to regulate the frequency and voltage in autonomous MGs is proposed in \cite{9039721}. A resilient controller for multi-agent systems under sensor and actuator attacks is presented in \cite{8439042}. The control follows the leader-follower scheme where the leader indicates the objectives that the follower agents should track. 
Similarly, two distributed control strategies utilizing a consensus-based leader-follower approach are demonstrated in \cite{sharma2017agent}. In both schemes, the agents' behavior is monitored for malicious actions and a linear quadratic regulator problem is defined to ensure the optimal power exchange between multiple storage resources. 
In \cite{ding2019distributed}, 
an acknowledgment-based detection algorithm is presented that serves as a preliminary step before recovering the compromised grid agents. The successful agent recovery is contingent upon the capability of the BESS to support power demand and meet the operation and transient constraints of power systems. 
Hence, an adapted secondary control scheme is proposed to maintain the frequency within acceptable limits.

The contribution of this work aims to leverage both aforementioned verticals, i.e., control and BESS resource allocation, and provide an automated and resilient methodology for the detection of maliciously operating DG agents in a MG setup. By isolating the affected DGs, their behavior can be detained safeguarding the operation of the rest of the grid. Then, leveraging the BESS resources, the power demand of the compromised (islanded DG) can be supplied while the MG system operator attempts to reactively restore the misbehaving agent to an trusted operating condition.

{The rest of the paper is structured as follows. Section \ref{s:systemModel} presents the cyber and physical system modeling of this work. Section \ref{s:Methodology} details the attack preliminaries and describes the cyber-attack mitigation methodology leveraging BESS resources, while Section \ref{s:results} provides the experimental evaluation of our approach. Last, Section \ref{s:Conclusion} concludes the paper and provides directions for future work.}
\section{Microgrid model and Control Objectives} \label{s:systemModel}

Centralized, decentralized, and distributed controls are  commonly used  in MGs. 
Although centralized control can monitor the entire system operation and realize accurate power sharing between converters, the communication burden limits its application to large-scale systems. Decentralized control alleviates the single point of failure 
and the communication overheads as it depends on  local information. In contrast,  distributed control reduces the overall communication dependency by utilizing local data from neighboring nodes. 

\subsection{Cyber Layer Modeling} \label{cyberLayer}

Distributed control schemes utilize peer-to-peer communication between  neighboring nodes via sparse distributed communication networks. 
The communication network can be seen as a topological map represented by an undirected graph $\mathcal{G}=(\mathrm{V}, \mathrm{E})$ with nodes $\mathrm{V}=\{1, \ldots, n\}$, and edges $\mathrm{E} \subseteq \mathrm{V} \times \mathrm{V}$. If $\mathrm{e}_{ij} = \mathrm{e}_{ji}= (i, j)\in  \mathrm{E} $, the nodes $i$ and $j$ are adjacent, thus can exchange information through the communication network. $\mathrm{A}=\left[a_{i j}\right] \in \mathbb{R}^{n \times n} $ is the 
adjacency matrix with non-negative adjacency elements, i.e., $a_{i j} = 0$, and $a_{i j}=1~\text{if and only if the edge}~(i, j) \in \mathrm{E}$. The neighbors of node $i$ are defined as $\mathcal{N}_{i} \triangleq\{j \in \mathrm{V}:(i, j) \in \mathrm{E}\}$.

\begin{figure}[t]
\centerline{\includegraphics[width=0.35\textwidth]{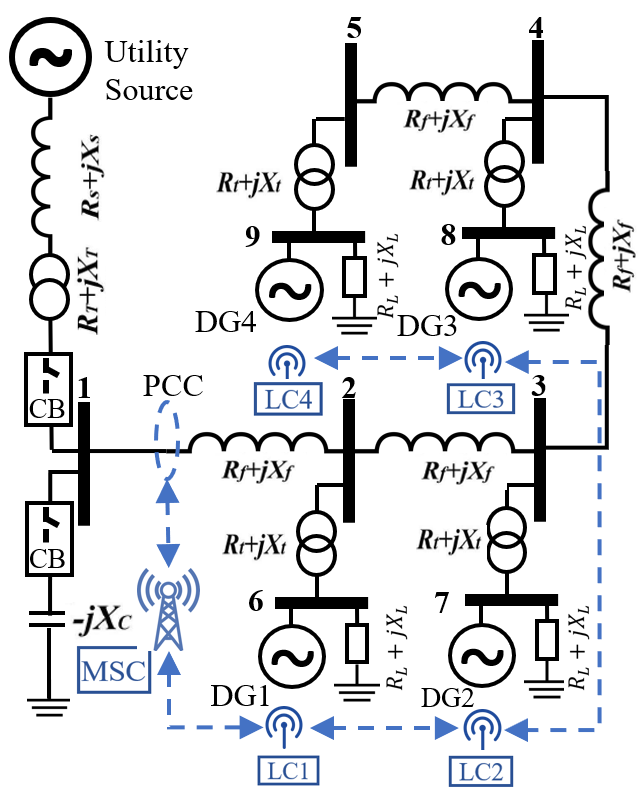}}
\caption{Structure of the studied microgrid: physical layer (black lines) and cyber layer (blue lines): Canadian urban benchmark distribution system.}
\label{fig:architecture}
\end{figure}

In this work, a modified version of the Canadian urban benchmark distribution system is utilized to demonstrate the leader-follower distributed control scheme, shown in Fig. \ref{fig:architecture} with blue color. It is considered that each converter has a local controller (LC), operating under such control. The data $\left \{ P, Q \right \}$ is exchanged between neighboring nodes to achieve primary control objectives, where $P,Q$ are the active and reactive power sharing values. For example, LC2 communicates and collects the data from LC1 and LC3, which can be represented as LC2 $=\left \{\xi_{12}, \xi_{32}\right \}$, where $\xi_{12} =\left \{{P_{1},Q_{1}}\right \}$ and $\xi_{32} =\left \{{P_{3},Q_{3}}\right \}$ are the data collected from LC1 and LC3 and transferred to LC2,  respectively. Furthermore, a mode supervisory controller (MSC) is placed close to the point of common coupling (PCC), which can exchange measurements and multicast signals to the respective LCs, realizing secondary control objectives.


\subsection{Physical Layer Modeling}
DGs can be integrated into an AC MG through power converters. 
When the MG is connected to the main grid, known as the grid-tied mode, the converters perform as current-controlled sources. The MG voltage and frequency are mainly regulated by the main grid. Otherwise, when the MG is disconnected from the main grid, known as the grid-forming or islanded mode, it will be stabilized by the converters which can operate in voltage-source or current-source mode. The modeling of MGs operating in their autonomous mode is investigated in this work. Furthermore, we leverage BESS resources to overcome adverse deliberate scenarios. 
Droop control is widely utilized to share power among parallel-connected converters and improve system reliability by imitating the parallel operation behavior of synchronous generators. It is considered an independent or autonomous form of control since it maintains stability while eliminating the synchronous communication requirement between  grid converters \cite{lo2012decentralized}. 

In order to model the MG system, we consider $n$ nodes consisting of loads and DGs (e.g., renewable sources or energy storage systems). The complex power delivered to the MG from the converter on node $i$ can be calculated by $ S=V_{\mathrm{g}} I^{*}_{o i}$, where $V_\mathrm{g}\angle 0$ represents the grid voltage and $I_{o}$ is the current injected into the grid. Thus, the converter active and reactive power output can be expressed as Eq. \eqref{eq:pq}, where $V_{{i}}\angle \beta_{i}$ is the converter output voltage and $Z_{i}\angle \theta_{i}$ represents the output filter and line lumped impedance between the converter and grid.
\begin{align}
\left\{\begin{array}{l}P_{G i}=\frac{V_{\mathrm{g}}V_{\mathrm{i}}}{Z_{i}} \cos (\theta_{i}-\beta_{i})-\frac{V_{\mathrm{g}}^{2}}{Z_{i}} \cos (\theta_{i}) \\ Q_{G i}=\frac{V_{\mathrm{g}} V_{\mathrm{i}}}{Z_{i}} \sin (\theta_{i}-\beta_{i})-\frac{V_{\mathrm{g}}^{2}}{Z_{i}} \sin (\theta_{i})
\end{array}\right.
\label{eq:pq}
\end{align}



When the phase difference is small enough and impedance $Z\angle \theta$ is assumed to be purely inductive with $\theta=90^{\circ}$, the $P-\omega$ and $Q-V$ droop control characteristic,  {also known as primary control,} can be described by Eq. \eqref{eq:Droop1},  where $\omega$ and  $V$ are the angular frequency and converter output voltage of the inverter, respectively, with $\omega_{0}$ and $V_{0}$ when there is no load, and $k_{P}$ and $k_{Q}$ are the droop coefficients. 
\begin{align}
\begin{aligned}
\omega = \omega_{0} - k_{P} \cdot P_{G} \\
V = V_{0} - k_{Q} \cdot Q_{G} 
\end{aligned}
\label{eq:Droop1}
\end{align}

Although the droop control can achieve independent active and reactive power control in the primary level or converter level without intercommunication between nodes, secondary level or system level control can assist in compensating the frequency and voltage deviation caused by the primary level droop control. {In the distributed control scheme under investigation, the secondary level control takes the voltage ($V_{s}$) and frequency ($\omega_{s}$) of the PCC point. Ultimately, the droop control can be formulated as Eq. \eqref{eq:Droop2}, where $\Delta \omega $ and $\Delta V$ are the secondary frequency and voltage contribution, respectively.} 
\begin{align}
\begin{aligned}
\omega = \omega_{0} - k_{P} \cdot P_{G} +\Delta \omega \\
V = V_{0} - k_{Q} \cdot Q_{G} + \Delta V 
\end{aligned}
\label{eq:Droop2}
\end{align}

In terms of Eqs. \eqref{eq:pq} -- \eqref{eq:Droop2}, a converter can be linearized as the following state-space model in Eq. \eqref{eq:pgm}  with $a$ inputs, $b$ states, and $c$ outputs \cite{611275}. 
The state derivatives $ \dot{\mathbf{x}} \in \mathbb{R}^{b} $ are calculated using the state variables $\mathbf{x} \in \mathbb{R}^{b} $ at time $t$ and system control input $\mathbf{u} \in \mathbb{R}^{a}$. $\mathbf{y} \in \mathbb{R}^{c}$ is the system output (sensor measurements). $\mathrm{A} \in \mathbb{R}^{b\times b}$, $ \mathrm{B}\in \mathbb{R}^{b \times a}$, $ \mathrm{C}\in \mathbb{R}^{c \times b}$ are the state, control, and output matrices of the physical system
and $\mathbf{e}$ denotes the noise vector of sensed measurements. 
\begin{align}
\begin{aligned}
 \dot{\mathbf{x}}_{t} &= \mathrm{A} \mathbf{x}_{t} + \mathrm{B} \mathbf{u}_{t}\\
 \mathbf{y}_{t} &= \mathrm{C} \mathbf{x}_{t} + \mathbf{e}_{t}
\end{aligned}
\label{eq:pgm}
\end{align}
The state variable $\mathbf{x}_{t}$ and control input $\mathbf{u}_{t}$ are expressed as:
\begin{align}
\begin{aligned}
{\mathbf{x}}&=\left[ \beta, v_{od}, v_{oq}, i_{od}, i_{oq}, i_{d}, i_{q}, P_{G}, Q_{G}, \omega, V \right]^{T} \\
{\mathbf{u}}&=\left[{\omega_\mathrm{s}},{V_\mathrm{s}}\right]^{T}
\label{eq:xandu}
\end{aligned}
\end{align}
where $i_{\mathrm{od}}$, $i_{\mathrm{oq}}$, $v_{\mathrm{od}}$, $v_{\mathrm{oq}}$, $i_{\mathrm{d}}$, $i_{\mathrm{q}}$ are the $d$- and $q$-axis output currents, output voltages, and currents of the converter.

\section{BESS Mitigation Methodology} \label{s:Methodology}

In this section, we define adversarial capabilities and attack assumptions. We detail the formalization for the leader-follower distributed control scheme and provide constraints for the BESS nominal operation. Finally, we discuss the malicious agent detection methodology and designate the steps of our proposed BESS-assisted MG self-healing mechanism. 

\subsection{Adversary and Attack Model} \label{s:attackModel}
We consider an adversary capable of attacking the cyber/communication layer by sending incorrect measurements to deceive the surrounding nodes. \textcolor{black}{The actual measurement $\mathbf{y}$ -- forwarded to the $\mathcal{N}_{i}$ nodes -- changes to $\mathbf{y}^{a}$ once the vulnerability of the communication channel between these 
neighboring nodes is exploited, where $\alpha$ denotes an attacked channel.} Since the relationship between the received measurements ($\widetilde{\mathbf{y}}$) and the control signals ($\mathbf{u}$) can generally be expressed as in Eq. (\ref{eq:attacky}), where $\mathrm{H}$ is a feedback control gain \cite{zhou2020secure},  when
the controller takes the malicious measurement $\mathbf{y}^{a}$ instead of the true measurement $\mathbf{y}$, where $\widetilde{\mathbf{y}} = \mathbf{y}^{a}$, the control scheme is misled by the attacker.    
\begin{align}
    \mathbf{u}(t) = -\mathrm{H} \widetilde{\mathbf{y}}(t)
 \label{eq:attacky}   
\end{align}

To establish such attacks targeting measurement signals, several models can be utilized \cite{mohan2020comprehensive}.  
The attack measurement can be described by $\mathbf{y}^{a}_{t}$, where $\mathbf{y}_t$ is the true measurement, and $[t_{a}, t_{a'}]$ is the attack duration between $t_{a}$ and $ t_{a'}$.

\subsubsection{The scaling attack} 

This type of attack could increase or reduce $(1+\mathbf{a}_s)$ times the measurement from the actual value by using the scaling attack vector $\mathbf{a}_s \in \mathbb{R}$ as shown in Eq. \eqref{eq:attackca}. 
\begin{align}
\mathbf{y}^{a}_{t}=\left\{\begin{array}{ll}
\mathbf{y}_{t} & \text { for } \mathrm{t} \notin [t_{a}, t_{a'}] \\
\left(1+\mathbf{a}_{s}\right) \mathbf{y}_{t}& \text { for } \mathrm{t} \in [t_{a}, t_{a'}]
\end{array}\right.
\label{eq:attackca} 
\end{align}

\subsubsection{The additive attack}
It adds an additive attack vector $\mathbf{a}_a$ on the original measurement during the attack period.  
\begin{align}
\mathbf{y}^{a}_{t}=\left\{\begin{array}{ll}
\mathbf{y}_{t} & \text { for } \mathrm{t} \notin [t_{a}, t_{a'}] \\
\mathbf{y}_{t} + {\mathbf{a}_{a}} & \text { for } \mathrm{t} \in [t_{a}, t_{a'}]
\end{array}\right.
\label{eq:attackaa} 
\end{align}

\subsubsection{The ramping attack}
The attack vector $\mathbf{a}_r$ is multiplied by the time scaling factor as following: 
\begin{align}
\mathbf{y}^{a}_{t}=\left\{\begin{array}{ll}
\mathbf{y}_{t} & \text { for } \mathrm{t} \notin [t_{a}, t_{a'}] \\
\mathbf{y}_{t} + {\mathbf{a}_{r}} t & \text { for } \mathrm{t} \in [t_{a}, t_{a'}]
\end{array}\right.
\label{eq:attackra} 
\end{align}

We assume that there are two ways to inject the signal of such attacks. The constant signal is considered a one-time attack, where the attacker implements the attack in a specific instant (e.g., $[t_{a}, t_{a'}]$). Another is called a periodic signal, where the attacker repeatedly injects malicious vectors within a period $T$ (e.g., $ [T+t_{a}, T+ t_{a'}]$). Specifically, when information is transferred between nodes through the cyber layer, the active and reactive power of neighboring converters can be manipulated. The primary droop control scheme can be deceived if malicious measurements are considered. Therefore, the system frequency and voltage stability are disturbed directly via droop control with  incorrect (manipulated) measurements.

\begin{figure}[t]
\centering
    \subfloat[]{
        \includegraphics[width=0.31\linewidth]{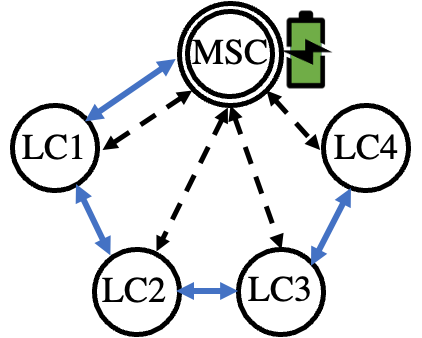}
        \label{fig:healthy}
    } 
    \subfloat[]{
        \includegraphics[width=0.31\linewidth]{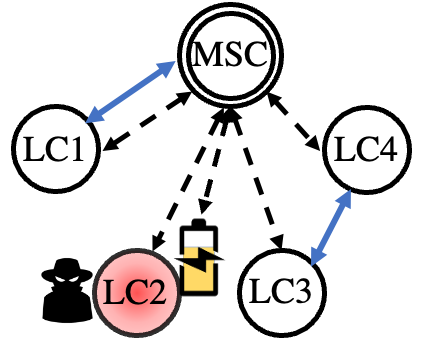}
        \label{fig:attacked}
    }
    \subfloat[]{
        \includegraphics[width=0.31\linewidth]{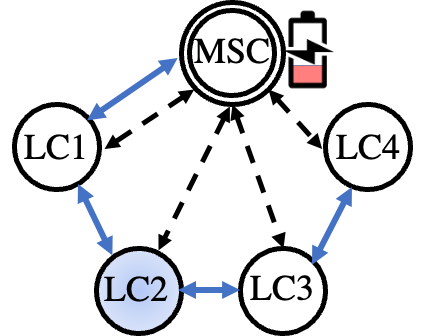}
        \label{fig:restored}
    }
\caption[CR]{Leader-follower distributed control scheme between DG agents and MSC. 
\subref{fig:healthy} Presents the healthy state of the MG, \subref{fig:attacked} demonstrates the isolation (communication-wise) of a compromised agent and the use of BESS to support local power demand, and \subref{fig:restored} illustrates the MG state after the attacked agent has been restored.} 
\label{fig:agents}
\end{figure}

\subsection{Distributed Control for Agent Systems} \label{distributed_control}

Leveraging the cyber modeling (Section \ref{cyberLayer}), we opted for a leader-follower consensus scheme to harden the control of MG agents in the presence of adversaries  \cite{sharma2017agent}. \textcolor{black}{Fig. \ref{fig:agents} depicts the communication architecture of the leader-follower scheme under nominal operation (Fig. \ref{fig:healthy}), unexpected conditions (Fig. \ref{fig:attacked}), where the malicious agent is removed from the distributed consensus set, and after agent restoration (Fig. \ref{fig:restored}).} In Fig. \ref{fig:agents}, the blue arrows represent the communication links between neighboring agents (exchanging information for their distributed primary control objectives), while the dashed arrows illustrate the information exchanged between MSC and DG agents (e.g., secondary control objectives, BESS coordination, MG restoration commands, etc.). \textcolor{black}{Access to every agent is maintained at all times by the MSC leader (dashed arrows).}  
In such a distributed control system, the agents' (i.e., DGs) objective is to track the state of the leader. The control goal at time $t$ for the $i^{th}$ agent is described by Eq. \eqref{eq:DCA} \cite{sharma2017agent}.
\begin{align}
\begin{aligned}
    u_{i}(t)=-h_{i}(t) \sum_{j=1, i\neq j}^{n} & \alpha_{ij}(t) [ (x_{i}(t) - x_{j}(t)) \\
    &- h_{i}(t)b_{i}(x_{i}(t) - x_{0}(t)) ] \\
\end{aligned}
\label{eq:DCA}
\end{align}
where $x_{i}(t)$, $ u_{i}(t)$, and $\alpha_{ij}(t)$ are the state information, the control input, and the element of the adjacency matrix $\mathrm{A}$ for the $i^{th}$ agent at time $t$. Agent $0$ is the leader and its corresponding state that the rest of the agents need to track is $x_{0}(t)$.  $h_{i}(t)$ is the local control gain and $b_{i}$ is equal to $1$ if the control information originates from the leader, else it is $0$. In leader-follower distributed control schemes, the DG LCs are responsible to track the operational system state as indicated by the leader. As a result, to attain consensus between all system nodes Eq. \eqref{eq:consensus} needs to be satisfied.
\begin{align}
\begin{aligned}
\label{eq:consensus}
    \lim_{t \to \infty}  |x_{i}(t) - x_{0}(t)| = 0
\end{aligned}
\end{align}

\subsection{Microgrid Recovery Mechanism using Energy Storage}
In our case, the leader of the system is  MSC. The MSC monitors and supervises the nominal operation of the system by issuing control commands to the DG agents, enforcing the economic operation, managing the BESS resources (at the PCC), and providing remediation actions in case of adverse events. For instance, after the detection of a misbehaving agent \cite{zografopoulos2021detection}, the leader will issue a disconnect command to the circuit breaker (CB) of that agent and at the same time instruct the remaining DG agents to remove the node from their consensus set. Recovery of the compromised agent will then be initiated with coordination between the affected node and the MSC. Leveraging the BESS reserves, the MSC will attempt the recovery of the maloperating DGs. The process is performed according to the secondary frequency control mechanism (Eq. \eqref{eq:Droop2}). The automatic recovery of the abnormally operating DGs can be performed 
granted that Eqs. \eqref{eq:bessPow} -- \eqref{eq:bessFreq} are satisfied. \textcolor{black}{However, if BESS storage cannot meet the power demand, MSC will enforce incremental load shedding to maintain critical loads online until the agent recovery.}


\begin{align}
\label{eq:bessPow}
   P_{BESS}^{min} \leq P_{BESS}(t) \leq P_{BESS}^{max} \\
\label{eq:bessSOC}
    SOC_{limit}^{min} \leq SOC_{BESS}(t) \leq SOC_{limit}^{max} \; , \forall \; t
\end{align}
where $P_{BESS}^{min}$ and $ P_{BESS}^{max} $ are the minimum and maximum limits of power that the BESS can supply, respectively, and $SOC_{limit}^{min}$ and $SOC_{limit}^{max}$ describe the state-of-charge (SOC) range within which the BESS should be operated. 
\begin{align}
\begin{aligned}
\label{eq:bessFreq}
    \lim_{t \to t+t_{lim}} = |\omega_{BESS}(t) - \omega_{ref}| = 0  \\
    \omega_{BESS}(t) = \omega_{ref} \; , \;
    \forall \; t \geq t+t_{lim}
\end{aligned}
\end{align}
Furthermore, the frequency of the DG when supported by BESS ($\omega_{BESS}(t)$) should be regulated to the reference frequency $\omega_{ref}$ within the system settling timing constraint $t_{lim}$.

\subsection{Malicious Agent Behavior Mitigation }

In the event of a cyberattack or an arbitrary adverse fault condition, one or more agents might start operating erroneously. Fault conditions are efficiently detected as they violate the system semantics in a congruent way, and automated mitigations to overcome such diverse effects are built-in within critical systems such as power grids. On the other hand, a cyberattack will manipulate an agent's operation in an ``indistinguishable from normal'' manner, by poisoning the information it shares with other agents, causing communication delays, blocking the data exchanged between agents, corrupting the control inputs to the underlying LC devices (within the DG), etc. By propagating the compromised information to the rest of the agents, the distributed control scheme can also be corrupted, hence jeopardizing the efficient operation of the whole MG. For instance, the state information of the compromised $i^{th}$ agent can be described using Eq. \eqref{eq:DCA}.
\begin{align}
\begin{aligned}
\label{eq:state}
    x_{i}(t+1) = x_{i}(t) + u_{i}(t)
\end{aligned}
\end{align}
where $ u_{i}(t)$ follows Eq. \eqref{eq:attacky}, and includes the attacked measurement components under attack models of Eqs. \eqref{eq:attackca} -- \eqref{eq:attackra}.

It is crucial for the MG system operation to detect and mitigate the effects of maliciously operating agents before their propagation to the system. For the detection of such conditions, different approaches have been pursued in the literature. Related work leverages acknowledgment-based messages exchanged between nodes that should arrive within predefined time limits to verify the nominal agent behavior \cite{ding2019distributed, sharma2017agent}. Other approaches utilize anomaly detection or reputation-based schemes to detect misbehaving agents \cite{zeng2014reputation}. 

Our methodology is based on the leader-follower control mechanism and leverages the information exchanged between DG agents and the MSC (e.g., system states) to identify suspicious behavior. Namely, conventional state estimation algorithms, targeted at transmission systems, cannot be applied to distribution systems as in our case. System state observability is limited in DGs due to the lack of temporal granularity, accuracy, and synchronization in the system measurements (contrary to the measurements provided by phasor measurement units in transmission systems). As a result, to improve the fidelity of the DG system models, synthetic or other data (from microPMUs, smart meters, DER assets, etc.) are used to enhance the accuracy of state estimation \cite{dehghanpour2018survey}.

Physics-informed methodologies considering the dynamics of power system assets, and their response (in attack-free cases and/or during different perturbations) have been explored in literature for the attack detection \cite{sahoo2019detection, zografopoulos2021detection}. In this work, we utilize a threshold-based approach where the cyber-physical properties of the system, i.e., state information $\mathbf{x}_{t}$ in Eq. \eqref{eq:xandu}, are supplemented with the DG inverter's controller dynamics, i.e., settling time, total harmonic distortion (THD), etc., and can signal abnormal operation. To maintain stable operation, acceptable levels for such properties in the attack-free case are defined by the system operators and device vendors \cite{5306080}. \textcolor{black}{The data collected during the DG operation account for measurement noise and other transient events that could occur during the system operation. As a result, custom-built system-specific models are developed for each DG (in the attack-free case). However, during malicious operations, including additive, scaling, or ramp attacks, the behavior of compromised agents can diverge from their expected values, as can be seen in Fig. \ref{fig:results}. Leveraging the custom-built data-driven DG models, cyberattacks can be distinguished from nominal behavior and/or other anomalous conditions (e.g., faults).}

Once abnormal agent behavior is identified, the MSC immediately issues an islanding command disconnecting the attacked DG from the system detaining the attack propagation to the rest of the DGs. \textcolor{black}{Furthermore, the local communication links between the compromised agent and its neighbors are disconnected, as can be seen in Fig. \ref{fig:attacked}, to limit the spread of untrustworthy false data (e.g., active, reactive power information). In the event that multiple DG agents are compromised simultaneously, the detection and isolation steps remain the same. However, the restoration process will differ since resources will have to be optimally allocated based on different criteria, e.g., load criticality, dispatchable BESS units, etc.} 

After the DG isolation, i.e., while operating in autonomous mode, the MSC will attempt the recovery of the compromised DG. The self-healing process can be performed either in coordination with the DG unit (if the controller has the ability to overcome the attack) or in a supervisory mode. In the latter case, which is presented in this work, the MSC (leader) will decouple the compromised unit (e.g., the inverter within the DG), and utilize the available BESS resources to supply the power demand of the DG. Granted that the BESS can support the corresponding demand (Eqs. \eqref{eq:bessPow}--\eqref{eq:bessFreq} are satisfied), the agent will be seamlessly reconnected with the rest of the system (plug-and-play capability). Once agent reconnection has succeeded, the MSC will attempt a black-start, that is, rebooting and restoring the DG inverter controller to a safe operational mode \cite{lopes2005control}. \textcolor{black}{Contrary to transmission systems, the radial nature of distribution systems, i.e., low connectivity index, allows the concurrent restoration of multiple DGs with minimal impact on the stability of the rest of the system nodes.} The algorithm delineating the steps of the BESS-assisted MG self-healing process is outlined in Alg. \ref{alg:healing}.

\begin{algorithm}[t]
\setstretch{1}
\small
\SetAlgoLined
\DontPrintSemicolon
\textbf{Input} DG agent \emph{Data}, BESS \emph{state} \\
\textbf{\#define} int {\it $\tau$}   \\
initialize.MG(read.Agent(\emph{Data})) \\
 \While{ MG.MSC(enabled) }{
     \For {a \textbf{in} \emph{Agents} }
        {
          \If{a.stateInfo() $\geq$ {\it $\tau$} }{ 
                \emph{\# Attacked agent detected}\\
                a.CircuitBreaker.disconnect() \\ 
                a.status = OFF \\
                
                \If{BESS.state() $==$ valid  }{ 
                    \emph{\# Support demand via BESS } \\
                    a.InverterController.disconnect() \\
                    BESS.supportDemand() \\
                    a.CircuitBreaker.connect() \\
                    a.InverterController.reboot()
                }
          }
        } 
        \emph{\# Rejuvenate compromised agents}\\
        \For {a \textbf{in} \emph{Agents.status $==$ OFF} }
        {
          \If{a.stateInfo() $\leq$ {\it $\tau$}  }{ 
                a.status = ON \Comment{Regenerate agent} \\
                a.InverterController.connect() \\
                BESS.disconnect() \\
          }
        } 
}
\caption{MG self-healing mechanism.}
\label{alg:healing}
\end{algorithm}

\section{Simulation Results} \label{s:results}

\begin{figure*}[t]
\centering
    \subfloat[]{
        \includegraphics[width=0.32\textwidth]{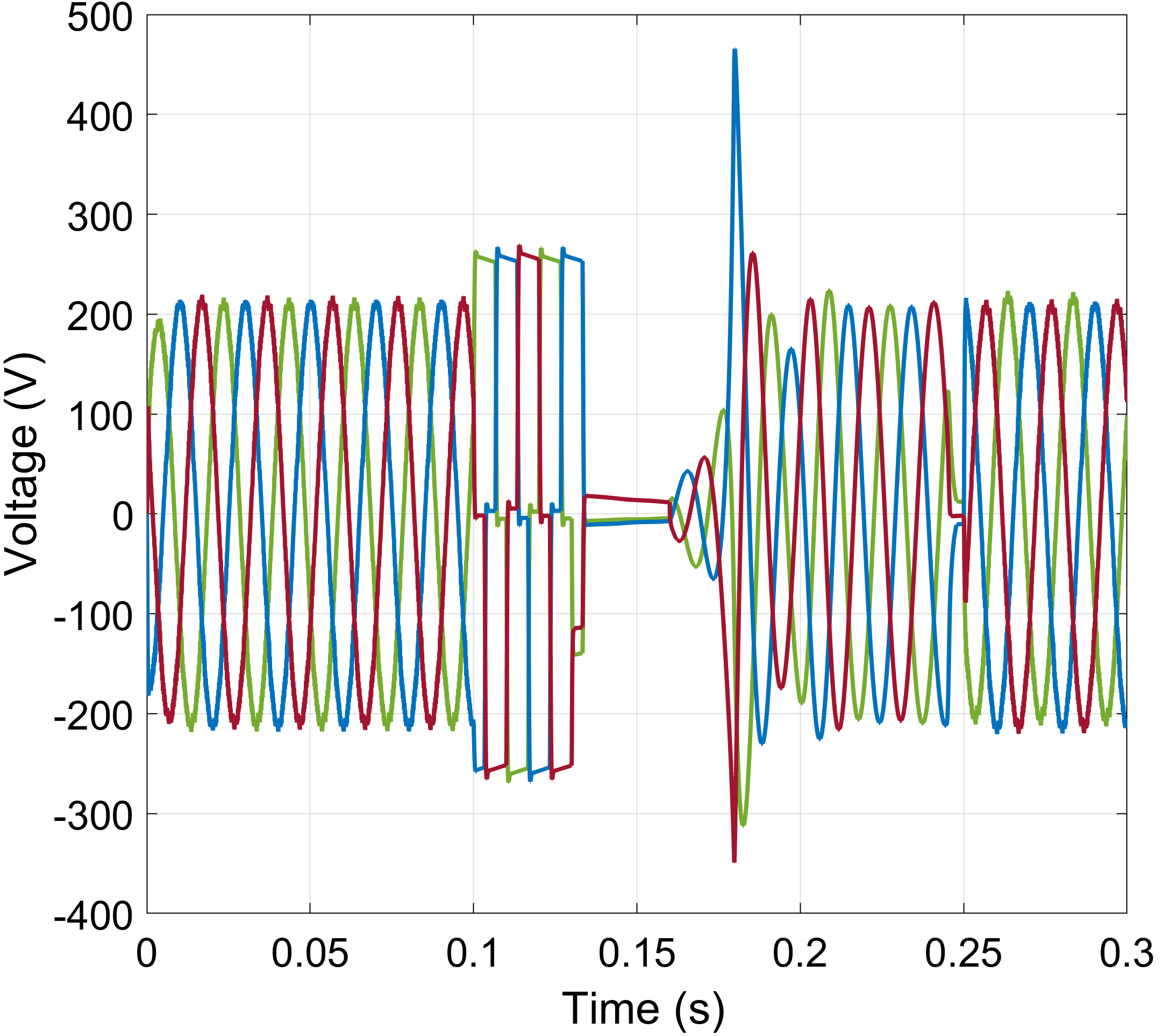}
        \label{fig:scaling}
    } 
    \subfloat[]{
        \includegraphics[width=0.32\textwidth]{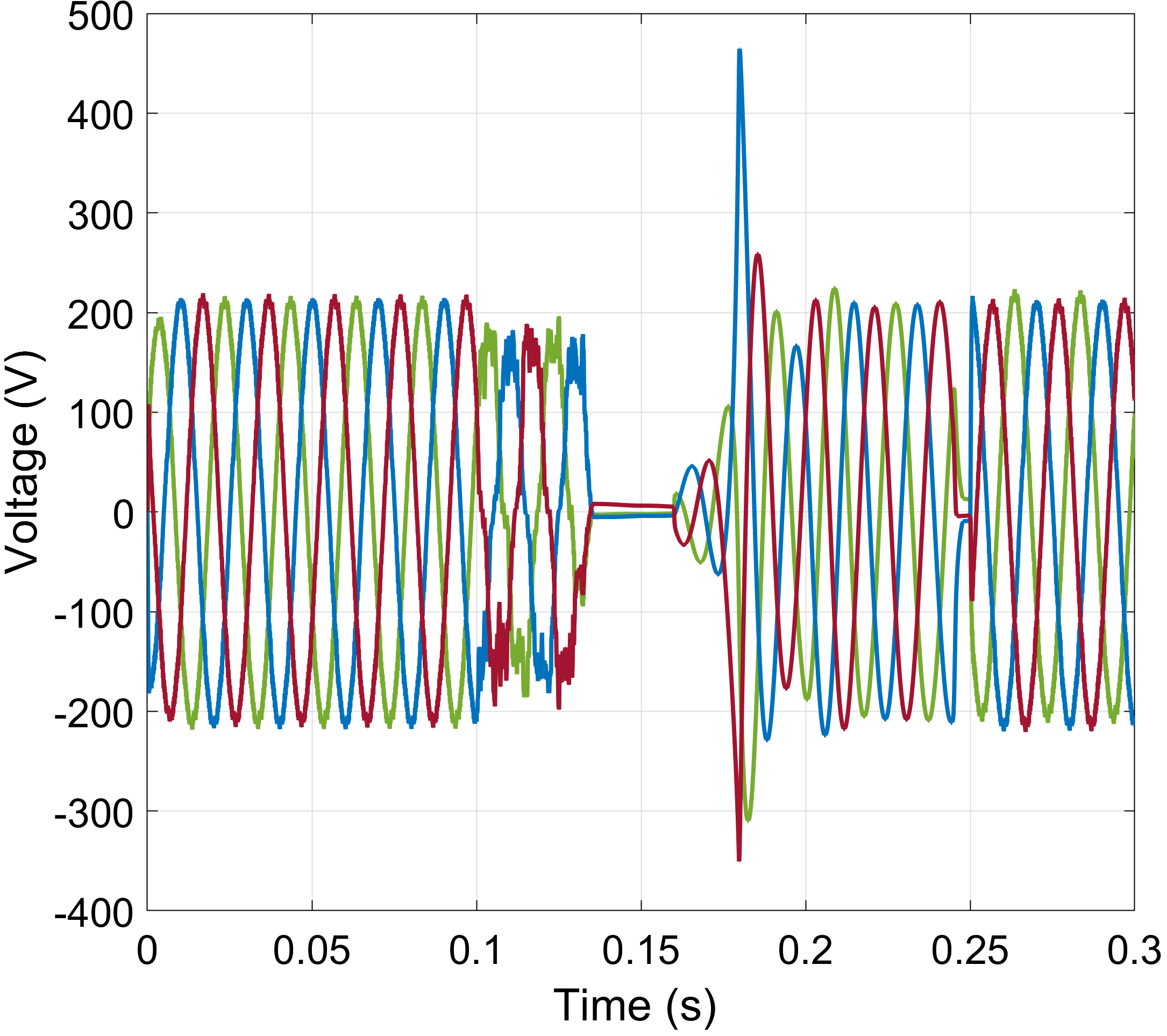}
        \label{fig:additive}
    }
    \subfloat[]{
        \includegraphics[width=0.32\textwidth]{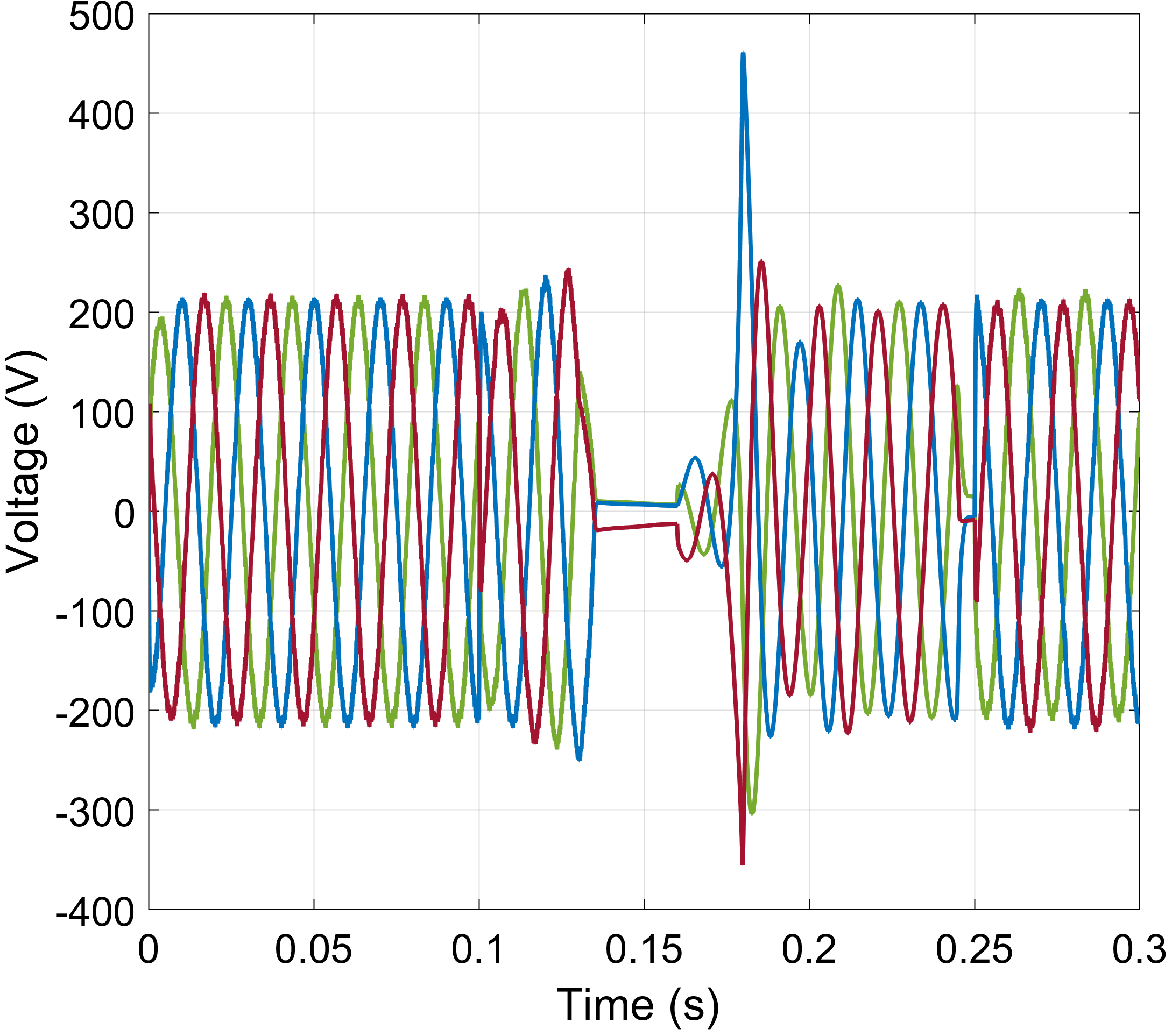}
        \label{fig:ramp}
    }
\caption[CR]{DG voltages during three different types of control input attacks: \subref{fig:scaling} scaling, \subref{fig:additive} additive, and \subref{fig:ramp} ramping attack. The attacks are initiated at $t=0.1s$, and the DG is disconnected from the MG at $t=0.14s$. At $t=0.16s$, the BESS is used to support the compromised DG power demand, and at $t=0.25s$, after the attacked agent has been recovered, the BESS hands over to the restored DG converter. 
} 
\label{fig:results}
\end{figure*}

For the evaluation of our BESS-assisted cyberattack mitigation strategy, we use \textsl{MATLAB's Simscape Electrical} toolkit. We simulate a modified version of the Canadian urban distribution system (depicted in Fig. \ref{fig:architecture}) that has been adjusted to include a BESS. The MG (bus 1) is connected through a CB to the utility grid via a 120kV/12.5kV distribution substation transformer. A 2.75MVar capacitor bank responsible for harmonic frequency suppression is also connected to bus 1 via a CB. The remaining MG buses support aggregated residential loads that operate with a constant power demand of 2 MW. The BESS, furnishing a capacity of $1$MWh, is placed at the PCC and is controlled by the MSC. It is interfaced with the MG feeder similarly to the DG nodes, i.e., via a $12.5$kV/$208$V step-down transformer. More details on the Canadian urban benchmark model can be found in \cite{zografopoulos2021detection}.

We have implemented three attack scenarios targeting primary control objectives of the DGs converters. Following the models described in Section \ref{s:attackModel}, we perform a scaling, an additive, and a ramping attack targeting the converter modulation controller. In Fig. \ref{fig:results}, we demonstrate the impact of these attack types on the three-phase voltages of the DG. In the scaling attack case (Fig. \ref{fig:scaling}), we observe that the converter's controller is saturated, thus producing an almost square output voltage waveform, while also exceeding the acceptable operational voltage limits (overvoltage condition). In the additive attack scenario (Fig. \ref{fig:additive}), we can observe that the attack is initiated at $t=0.1s$ and manifests itself as a superposition of a random noise signal on the three-phase voltage envelopes. In the ramping attack scenario (Fig. \ref{fig:ramp}), the output voltage follows an arbitrarily increasing pattern. Once anomalous agent behavior is detected ($t=0.14s$), the MSC issues a sectionalization command to the compromised DG's CB, and at $t=0.16s$ the BESS is employed to support the DG power demand. While the compromised DG is supported by the BESS, the MSC attempts to recover the attacked agent (similar to black-start procedures). \textcolor{black}{The DG controller restoration process is outside the scope of this work. Agent restoration could be performed by rolling back to a trusted/trustworthy firmware version, operation of the agent using only trusted functions (hardware-isolated security enclaves), secure updates, etc.} Once the DG agent is restored in coordination with the MSC, the BESS-to-converter handover occurs.  In our experiment, this handover takes place at $t=0.25s$, when we can observe a small transient perturbation, mainly attributed to the plug-and-play features of the BESS. 

\begin{figure}[t]
\centering
    \subfloat[]{
        \includegraphics[width=0.48\linewidth]{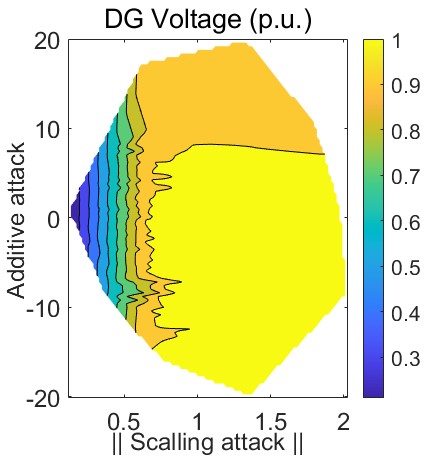}
        \label{fig:voltage}
    } 
    \subfloat[]{
        \includegraphics[width=0.49\linewidth]{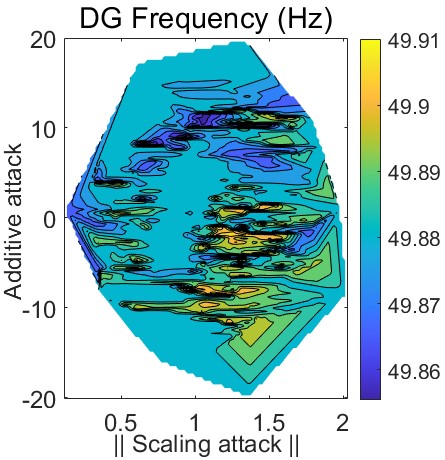}
        \label{fig:frequency}
    }

\caption[CR]{Steady-state attack impact 
of additive and scaling control input attacks on the DG \subref{fig:voltage} voltage, and \subref{fig:frequency} frequency.  } 
\label{fig:sensitivity}
\end{figure}

Sensitivity analysis methodologies can be used to identify the most critical components and how they affect system behavior. Monte Carlo simulations exhaustively search for such components or configurations and assist in prioritizing and mitigating undesirable conditions \cite{milanovic2017probabilistic}. In our case, sensitivity analysis is used to identify which additive, scaling, or ramping factors can maximize the impact of cyberattacks. In Fig. \ref{fig:sensitivity}, we demonstrate a heat map of the steady-state impact of control input attacks on the voltage and frequency of the DG, respectively. The vertical axis represents the additive attack magnitudes, whereas the horizontal axis accounts for scaling perturbations. \textcolor{black}{The attack-free scenario occurs when the additive attack magnitude (y-axis) is $0$ and the scaling attack factor (x-axis) is equal to $1$. Deviating from this operating point, 
the steady-state impact of additive and scaling attack combinations can be observed in Fig. \ref{fig:sensitivity}.} Ramping attacks can be abstracted as a combination of additive and scaling attacks; therefore, their steady-state impact is considered in Fig. \ref{fig:sensitivity}. The Monte Carlo-based sensitivity analysis illustrated in Fig. \ref{fig:sensitivity}, also indicates which composite additive and scaling attacks should be subverted, since they would maximize adversarial objectives and degrade system performance.

\section{Conclusions and Future Work} \label{s:Conclusion}

This paper proposes a BESS-assisted mitigation scheme that can maintain stable MG operation under unintentional or deliberate disturbances (e.g., faults, cyberattacks, etc.). The proposed mitigation methodology is two-fold. The first part is responsible for the detection of the attack and the isolation of the compromised agent. The second leverages the BESS to support the MG's power demand avoiding load shedding,  while attempting the automated restoration of the attacked agent and its IBR distributed generator. Our future work will focus on compound mitigation strategies which will ensure the optimal power dispatch leveraging synergies between BESS and other DERs during disruptive attack models. Furthermore, robust distributed misbehaving agent detectors in coordination with BESS planning will be employed to improve grid resilience in weakly connected MG systems.


\bibliographystyle{IEEEtran}
\bibliography{biblio}

\end{document}